\documentclass[traditabstract]{aa}

\usepackage{epsfig,graphicx,natbib}
\usepackage{longtable}
\usepackage{amssymb}
\usepackage{amsmath}
\usepackage{color}

\usepackage{bibentry}
\newcommand{\ignore}[1]{}
\newcommand{\nobibentry}[1]{{\let\nocite\ignore\bibentry{#1}}}

\def\PKS1830{PKS\,1830$-$211}

\begin{document}

\title{ALMA full polarization observations of \PKS1830\ during its record-breaking flare of 2019}

\author{  I. Marti-Vidal \inst{1,2}
  \and S.~Muller \inst{3}
  \and A.~Mus \inst{1,2}
  \and A. Marscher \inst{4}
  \and I. Agudo \inst{5}
  \and J. L. Gomez \inst{5}
}

\institute{
  Observatori Astron\`omic, Universitat de Val\`encia, Parc Cient\'ific, C. Catedr\`atico Jos\'e Beltr\'an 2, 46980 Paterna, Val\`encia, Spain
  \and
  Departament d’Astronomia i Astrof\'isica, Universitat de Val\`encia, C. Dr. Moliner 50, 46100 Burjassot ,Val\`encia, Spain
  \and  
  Department of Space, Earth and Environment, Chalmers University of Technology, Onsala Space Observatory, SE-43992 Onsala, Sweden
  \and
  Institute for Astrophysical Research, Boston University, 725 Commonwealth Avenue, Boston, MA 02215
  \and
  Instituto de Astrofísica de Andaluc\'ia, CSIC, Apartado 3004, 18080 Granada, Spain
}

\date {Received  / Accepted}

\titlerunning{Full polarization observations of \PKS1830}
\authorrunning{Marti-Vidal et al.}

\abstract{We report Atacama Large Millimeter Array (ALMA) Band 6 full-polarization observations of the lensed blazar \PKS1830\ during its record-breaking radio and gamma-ray flare in the spring of 2019. The observations were taken close to the peak of the gamma activity and show a clear difference in polarization state between the two time-delayed images. The leading image has a fractional polarization about three times lower than the trailing image, implying that significant depolarization occurred during the flare. In addition, we observe clear intra-hour variability of the polarization properties between the two lensed images, with a quasi-linear increase in the differential electric-vector position angle at a rate of about two degrees per hour, associated with changes in the relative fractional polarization of $\sim$10\%. This variability, combined with the lower polarization close to the peak of gamma activity, is in agreement with models of magnetic turbulence to explain polarization variability in blazar jets. Finally, the comparison of results from the full and differential polarization analysis confirms that the differential polarization technique (\citealt{mar16}) can provide useful information on the polarization state of sources like gravitationally lensed radio-loud quasars.}

\keywords{polarization -- quasars: individual: \PKS1830 -- Gamma rays: general}
\maketitle

\section{Introduction}

Relativistic jets from active galactic nuclei (AGN) reveal the powerful mechanisms at work in these extreme objects. Magnetic fields are expected to have a determinant impact in the launching and shaping of these jets, and it is therefore important to study their structure and evolution at different spatial and temporal scales. The magnetic fields can be probed by polarized emission (intensity and electric-vector position angle, EVPA), and the EVPA variability and correlations among different regions of the electromagnetic spectrum (from radio to $\gamma$ rays) bring strong observational constraints. Changes in AGN EVPA have been observed on timescales from days \citep[e.g.,][]{marscher2010} to months \citep[e.g.,][]{Agudo2018}, and from optical \citep[e.g.,][]{kikuchi1988} to radio \citep[e.g.,][]{MarsNat}. There is evidence of EVPA correlation between optical and radio, especially toward high radio frequencies (i.e., at millimeter and submillimeter wavelengths; e.g., \citealt{Algaba}). Correlation between strong EVPA changes and $\gamma$-ray activity have also been reported \citep[e.g.,][]{MarsNat} and the connection between radio and $\gamma$ rays is the subject of extensive observational and theoretical studies \citep[e.g.,][]{R1,mar13,R2}.

There are different models to explain the EVPA variability in blazars and its correlation with $\gamma$ rays, such as smooth (e.g., helical) magnetic-field structures \citep[e.g.,][]{Mys2018} coupled to geometric (e.g., Doppler boosting induced by jet precession) or magneto-hydrodynamic (e.g., shock-compression) effects. Alternatively, pseudo-random variability induced by turbulence can happen downstream in the jet in regions of $\gamma$-ray up-scattering \citep[for descriptions of some models, see, e.g.,][]{MarsNat,Zhang,Mars2014,Abdo2010}. The key mechanisms for the polarization variability may be related to the innermost regions of the AGN jets, which can only be probed with observations at the highest radio frequencies (i.e., millimeter and submillimeter wavelengths), due to effects of synchrotron self-absorption \citep[e.g.,][]{Blandford}.

A key observable to distinguish between the different models is the statistics of the EVPA variability on short timescales since models inferring turbulence predict stronger and more erratic variability \citep[e.g.,][]{Mars2014,Kiehlmann}. Unfortunately, millimeter and submillimeter polarization observations of AGN have been limited thus far by sensitivity and polarization accuracy, especially for measurements on short timescales. 

One of the best sources to detect intra-hour (sub)millimeter polarimetry variability is the gravitationally lensed blazar \PKS1830.
The source has two bright and compact images (hereafter labelled NE and SW), separated by 1$''$ and delayed in time by $\sim$26~days (see, e.g., \citealt{lov98}, NE image leading). Strong polarization has been reported at centimeter and millimeter wavelengths \citep[][]{sub90,gar98}. The submillimeter variability and polarization of \PKS1830\ have recently been studied with ALMA on timescales down to intra-day \citep{mar13,mar15,mar19}, and the Faraday rotation could even be estimated at submillimeter wavelengths \citep{mar15}. In addition, \PKS1830\ is a strong $\gamma$-ray emitter, with several flaring events reported in recent years by the Fermi-LAT collaboration \citep[e.g.,][]{AbdoPKS}, which makes it an interesting target to investigate the radio-to-$\gamma$-ray connection.

In March 2019, \PKS1830\ experienced a period of intense activity, with a record-breaking flare in the radio and $\gamma$-ray (cf., ATel \#12594, \#12601, \#12603, \#12622, \#12685, \#12739), which lasted several weeks. Here we report full polarization radio-millimeter observations of \PKS1830\ obtained with ALMA close in time to this exceptional peak of the $\gamma$-ray activity.

\section{Observations}

\PKS1830 was observed with ALMA in full polarization mode on 2019 April 11. Observations were done in two consecutive executions over a time span of about two hours. The Band~6 receivers were used and the correlator was configured to cover four independent 1.875~GHz wide spectral windows, centered at $\sim 234$, 236, 247, and 249~GHz, with a spectral resolution of 1.95~MHz per channel. 

The data calibration was done within the Common Astronomy Software Applications (CASA\footnote{http://casa.nrao.edu/}) package, following a two-step procedure. The first step was a standard calibration of the data. The bandpass response of the antennas was estimated from observations of the bright quasar J\,2000$-$1748 in the first execution and the solutions were used to correct the data from the two executions. The gain calibration was performed on the quasar J\,1832$-$2039, which was observed regularly every 5--7~min. The quasar J2000-1748 was also used as absolute flux calibrator. The flux-scaling factor was bootstrapped to J\,1832$-$2039 to properly scale the two executions. We estimate a flux accuracy better than 10\%, based on the ALMA monitoring data of this source. At this step, we could already inspect the spectra and flag the main absorption features from the z=0.89 molecular absorber in front of \PKS1830\ \citep[e.g.,][]{mul14} and from atmospheric lines, to finally produce (at the end of the full calibration process) pure continuum data by aggregating line-free channels.

The second part of the data reduction was dedicated to the full-polarization calibration, with estimate of the cross-polarization phases for the reference antenna and of the instrumental polarization of all antennas, following the standard ALMA calibration procedure \citep[e.g.,][]{Nagai}. The quasar J\,1924$-$2914 was observed as the polarization calibrator. For this purpose it was monitored regularly across the two executions to cover a wide range of parallactic angles.

After the polarization calibration, the visibility phases were self-calibrated, based on the continuum image of \PKS1830 in Stokes $I$, applying phase-only gain corrections at each 6\,s  integration and combining all four spectral windows to a common solution. This self-calibration step improved dramatically the quality of the data by reducing the atmospheric decoherence. In particular, the dynamic range (DR; i.e., the ratio of peak flux to rms noise level in the image) improved by a factor 2--3 after self-calibration, and reached values $\gtrsim 1000$ toward the high end expectations for ALMA observations\footnote{see, e.g., the ALMA Technical Handbook for Cycle 7, \S\,10.5.1.}. This is possible due to the high flux density of the two images of \PKS1830, which allow us to derive robust self-calibration gain corrections even with the short integration of 6~s.

\section{Results} \label{FullPolResSec}

The CASA task \texttt{tclean} was executed to reconstruct the images of \PKS1830 in Stokes I, Q, and U, separately. The CLEANing of Stokes Q and U was executed using the same mask as in the CLEANing of Stokes I. Uniform visibility weighting was used in the image deconvolution. The resulting beam is $0.47 ''\times 0.31''$, with a position angle  $-83^\circ$. The images of the degree of linear polarization, $m$, and EVPA, $\phi$, were computed pixel-wise from the Stokes Q and U images with the CASA task \texttt{immath}, following the well-known equations

\begin{equation}
m = \frac{\sqrt{U^2 + Q^2}}{I} ~~ \textrm{and} ~~ \phi = \frac{1}{2}\textrm{arctan}\left( \frac{U}{Q} \right).
\end{equation}

\begin{figure}[h] 
\begin{center}
\includegraphics[width=7.5cm]{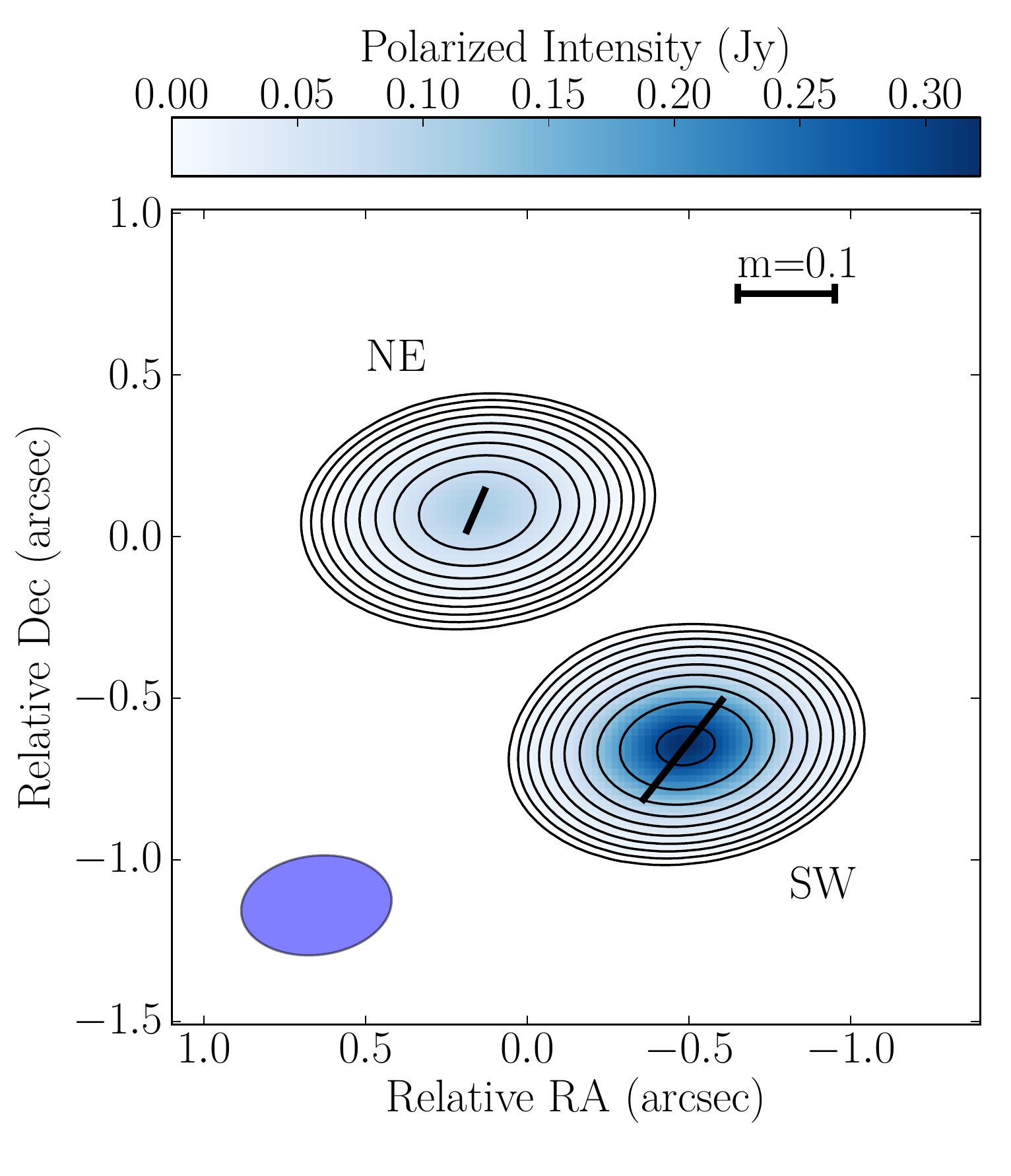}
\caption{ALMA band~6 full-polarization image of \PKS1830. Contours represent Stokes $I$ and are given in logarithmic scale, running from 2\% to 90\% of the peak intensity (2.34 Jy/beam); the  polarized intensity is color-coded (see scale at top); the thick lines represent the EVPA, with lengths proportional to fractional polarization (the peak of 14\% is located at the SW image). The scale of the fractional polarization, $m$, is given by the line in the top right corner; the full width at half maximum  of the CLEAN beam is shown in the bottom left corner.}
\label{fig:pksmap}
\end{center} \end{figure}

The final full-polarization image is shown in Fig.\,\ref{fig:pksmap}. The two lensed images show clear polarization signals, with a fractional polarization peak of 14\% (in the SW image, with a flux density of 2.34\,Jy) and 5.4\% (in the NE image, with a flux density of 2.10\,Jy). The most interesting result is the contrast in linear polarization between the lensed images. The leading NE image has a fractional polarization about three times lower than the SW image and its EVPA ($-24$\,degrees) is rotated $\sim$15 degrees with respect to that of the SW. This indicates that a significant evolution of the intrinsic polarization properties of \PKS1830\ occurred within a time interval of the order of the time delay (i.e., $\lesssim 26$~days).

\begin{figure}[h] 
\begin{center}
\includegraphics[width=8.5cm]{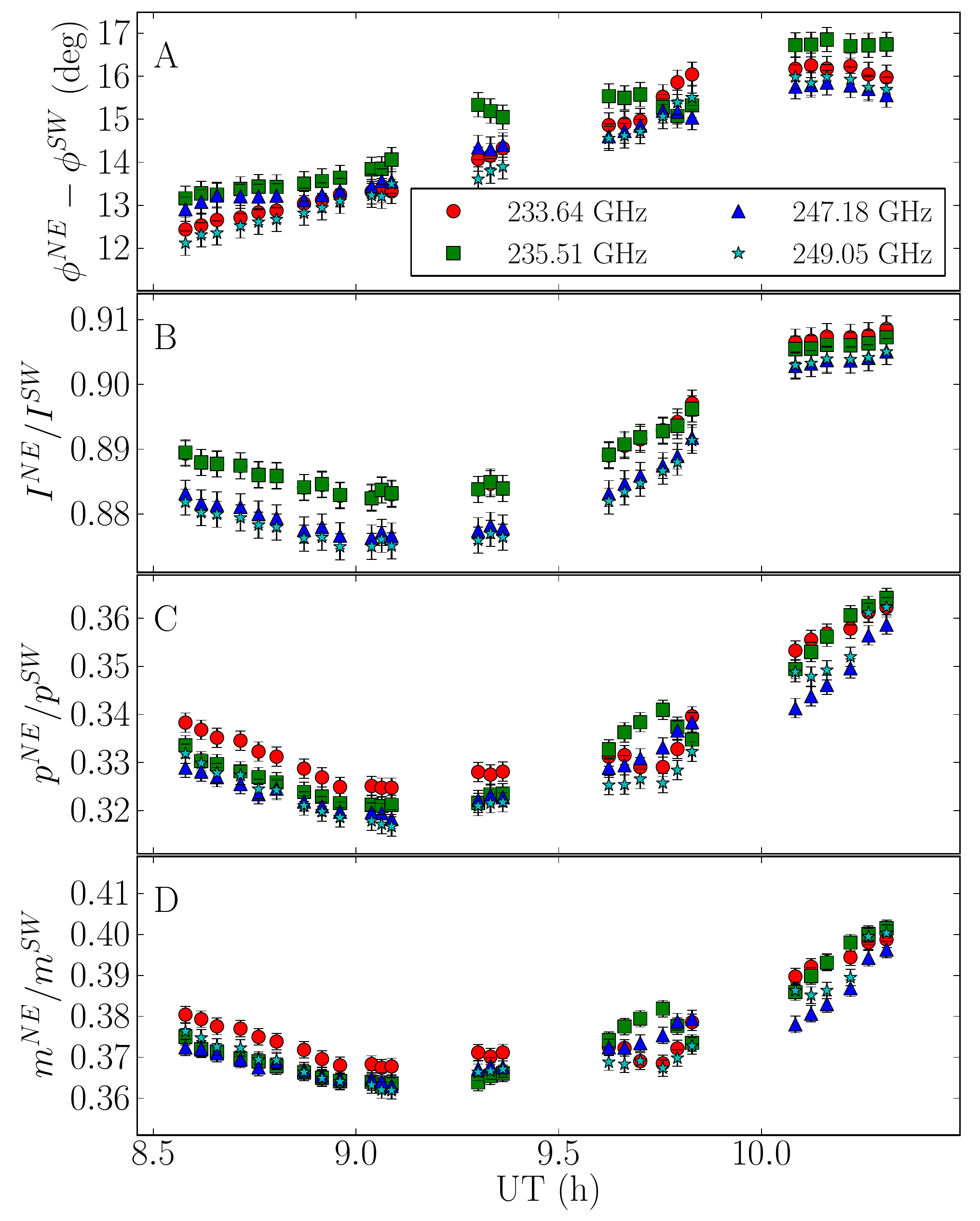}
\caption{Intra-observation evolution of the EVPA difference between the NE and SW images (A), ratio of total intensity of NE to SW (B), ratio of linear polarization intensity (C) and of fractional polarization (D). Each spectral window is shown with a different symbol and color.}
\label{fig:curves}
\end{center} \end{figure}

Because the data have high signal-to-noise ratio, it is also possible to search for flux and polarization variability within the time interval of the observations. Since both lensed images of the blazar are point-like within the ALMA beam, we can use direct visibility fitting with a simple and analytical source model;  among other advantages, this allows us to remove potential deconvolution problems. For a thorough discussion of the advantages of visibility model-fitting in this case, see \cite{mar14}. Each ALMA scan (with a typical duration of 410 seconds) was divided into three segments of equal length and the Stokes parameters I, Q, and U of \PKS1830 were fitted independently to each scan subdivision.  The fitting was performed with the Python-based task \texttt{uvmultifit} \citep{mar14} and the results are shown in Fig.\,\ref{fig:curves}. We note that our method uses a differential approach, which means that we search for relative changes between the NE and SW polarization states, in order to remove any possible residual instrumental effect that would affect the two images in the same way. We can therefore achieve the high accuracy required for the analysis of intra-observation variability (see, e.g., \citealt{mar13,mar14}).

The results show a clear linear increase in the EVPA difference between the two images (see Fig.\,\ref{fig:curves}A) at a rate of about two degrees per hour, and a hint of nonlinear behavior (a flattening in the time evolution) toward the end of the observations. The changes between spectral windows (which would encode a difference between the Faraday rotation in NE and SW) show an erratic trend, with no clear dependence of EVPA differences with wavelength squared. Even though a deep analysis of the differential NE/SW Faraday rotation from these ALMA data is beyond the scope of this paper, we present a short discussion of the EVPA wavelength dependence in Appendix~\ref{appendix:Faraday}.

Regarding the changes in polarization intensity (Fig.\,\ref{fig:curves}, panels C and D), we see a clear trend in the ratio between NE and SW polarization intensities (panel C) and fractional polarizations (panel D), which starts at around $m^{NE}/m^{SW} \sim 0.38$, reaches a minimum around $\sim0.36$ (half an hour later), and then rises monotonically up to $\sim 0.4$ at the end of the experiment (formal uncertainties on the order of $2\times10^{-3}$). There is a similar trend in the total-intensity ratio (Fig.\,\ref{fig:curves}B), which changes from $\sim 0.89$ to $\sim0.91$, with actual values slightly depending on frequency. Since the relative change in $I^{NE}/I^{SW}$ across the experiment is small (a 3.4\% difference between the maximum and minimum measured values, panel B) compared to the change in the polarization intensity ratio, $p^{NE}/p^{SW}$ (14.2\%, panel C), the time evolution of the relative fractional polarization (panel D) and polarization intensity (panel C) is similar. In addition, the correlation seen between Stokes I (panel B) and polarization intensity (panels C and D) may indicate that the source variability is dominated by the changes in polarized emission. To characterize the frequency dependence of the intensity ratio, $\rho$, we model it as a function of frequency and time using the expression 

\begin{equation}
\rho(t,\nu) = \frac{I_{NE}}{I_{SW}} = \rho_0(t) \times \left(\frac{\nu}{\nu_0}\right)^{\Delta \alpha(t)},  
\end{equation}

\noindent where $\Delta\alpha$ is the difference between the spectral indices of NE and SW. From the results shown in Fig.\,\ref{fig:curves} (B), we estimate a variation from $\Delta\alpha \sim -0.19$ to $\Delta\alpha \sim -0.05$ between the start and end of the observations. These results imply an intra-hour evolution of the blazar spectrum, although it is not possible to disentangle which of the two lensed images (or maybe both) is changing its spectral index.

\section{Discussion}

\subsection{Connection with $\gamma$-ray activity}

\begin{figure}[h] \begin{center}
\includegraphics[width=9cm]{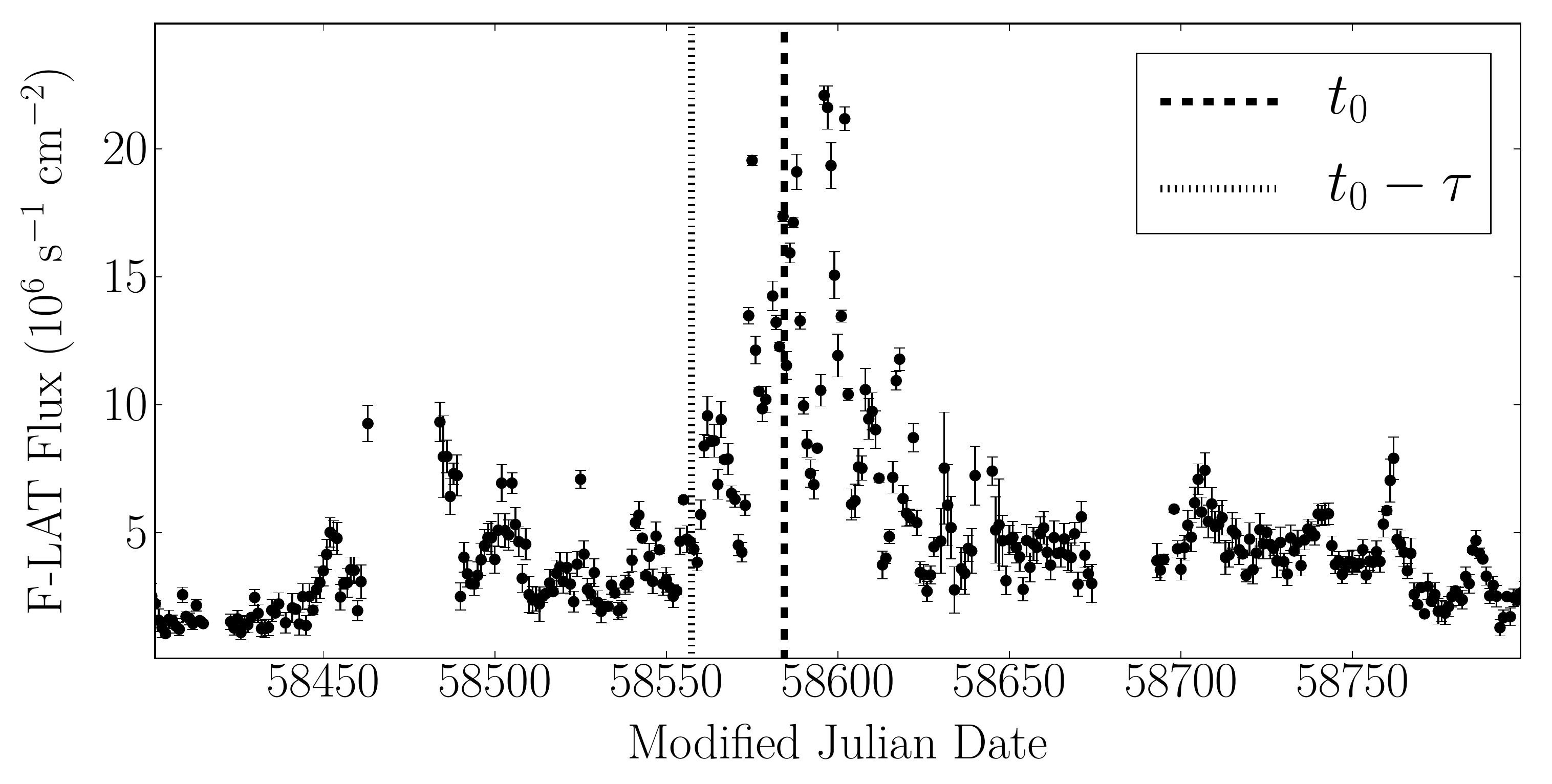}
\caption{Fermi-LAT $\gamma$-ray light curve, obtained from the LAT daily monitoring. The dashed line indicates the time of our ALMA observations, $t_0$, and the dotted line indicates the intrinsic time of the SW image (i.e., delayed by $\tau =26$~days). }
\label{fig:gamma-lc}
\end{center} \end{figure}

Given the time delay between the two lensed images of the blazar, we see the SW image in the same state as was the NE image $\sim 26$~days before the actual observations; or, alternatively, the NE image is seen in the same state as the SW image will be $\sim 26$~days after. Therefore, the lens properties allow us to probe the polarization of the blazar at two different epochs from one single ALMA observation (see the dotted and dashed lines in Fig.\,\ref{fig:gamma-lc}).

Our observations were taken about one month after the onset of a strong $\gamma$-ray flare in the lensed blazar, and about ten days before the $\gamma$-ray peak (see Fig.\,\ref{fig:gamma-lc}, generated from Fermi-LAT observations \footnote{See \texttt{https://fermi.gsfc.nasa.gov}}). This means that a comparison between the NE and SW images from our ALMA data is equivalent to a comparison between the blazar polarization states in the onset of the flare and about one month afterward (for the NE and SW image, respectively). As explained in Sect. \ref{FullPolResSec}, the fractional polarization of the leading image, NE, is about three times lower than that of the SW. Hence, a direct conclusion drawn from our full-polarization image is that the polarization intensity of the blazar was somehow suppressed during the $\gamma$-ray flare. In addition, the study of the EVPA difference between the NE and SW images implies that the polarization of the blazar during the $\gamma$-ray activity was varying notably on an intra-hour timescale. 

We can use the changes in the NE/SW polarization state to constrain the mechanisms of high-energy emission related to the flare. In particular, according to the turbulent model of blazar polarization variability \citep{Mars2014} the lower value of the fractional polarization, $m$, near the $\gamma$-ray peak may correspond to a stronger magnetic turbulence, with an increase in the number of turbulent cells of a factor of $\sim$9, since the number of cells would scale inversely as the square of the fractional polarization. The stronger turbulence could be related to enhanced particle acceleration (via magnetic reconnections or second-order Fermi acceleration) that caused the $\gamma$-ray flare, and would also be consistent with the rapid variability of both $m$ and the EVPA. In addition, the hint of nonlinear EVPA variation toward the end of our observations may also have a connection with either the ordered component of the magnetic field and/or the jet geometry (e.g., a changing pitch angle during the observations, \citealt{MarsNat}, and/or motions in a curved jet, \citealt{Mys2018}).

\subsection{Differential polarimetry}

Finally, we used the full-polarization data to validate the differential polarimetry technique used previously on several dual-polarization ALMA-based observations \citep{mar15,mar16,liu16,mar19}. This is the first time that we can directly compare the differential-polarimetry results to a full-polarization image, as a bona fide validation of the method. 

\begin{figure}[h] 
\begin{center}
\includegraphics[width=8.5cm]{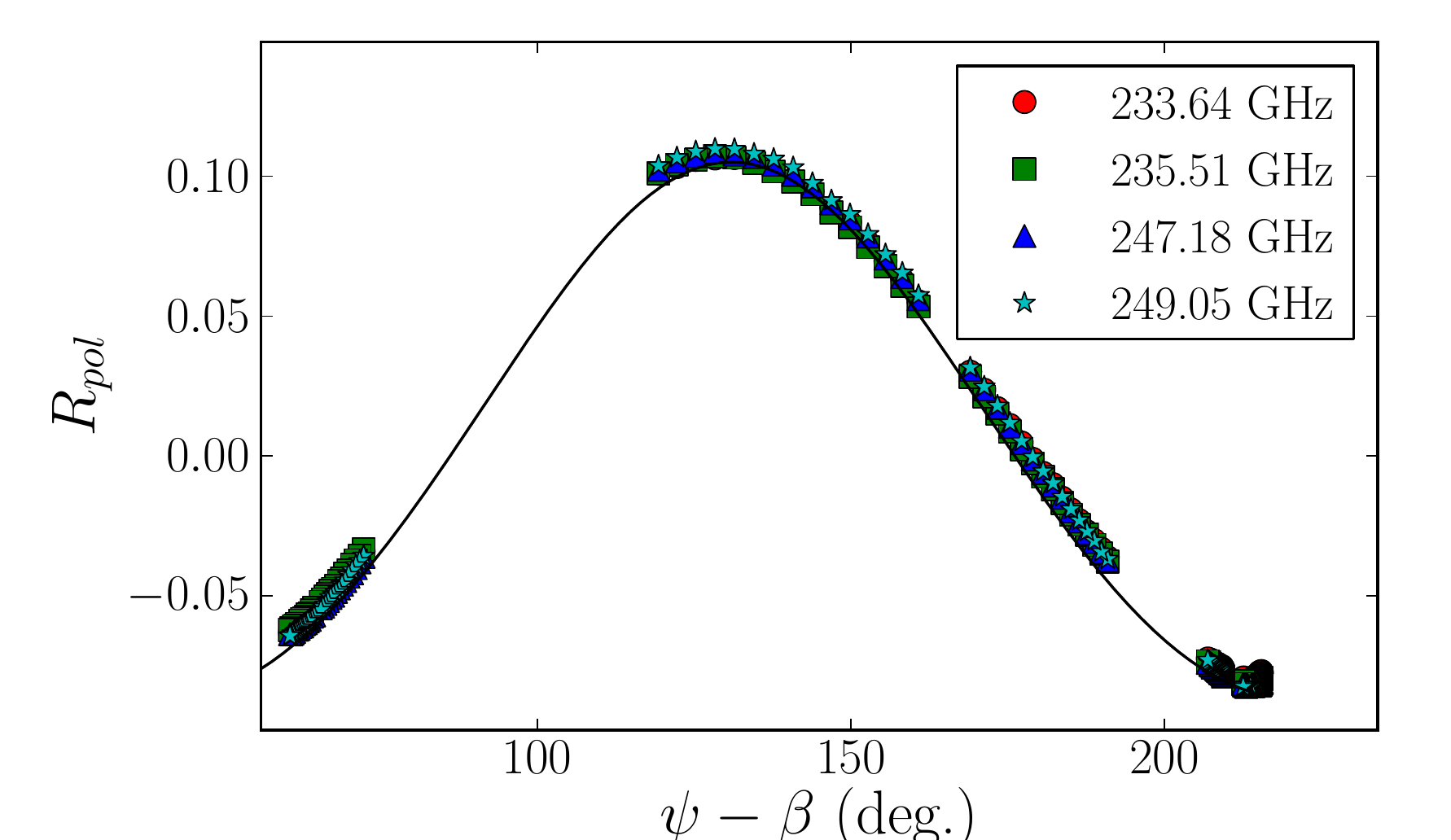}
\caption{Differential polarimetry, $R_{pol}$, between NE and SW as a function of the parallactic angle, $\psi$, rotated to the antenna feed, $\beta$ \citep[see Eqs. 13 to 16 in][]{mar16}. The thick line is the $R_{pol}$ prediction from the full-polarization image (i.e., the image shown in Fig.\,\ref{fig:pksmap}), assuming no Faraday rotation. Notice that the $R_{pol}$ values are similar for all frequencies.}
\end{center}
\label{fig:diffpol}
\end{figure}

Accordingly, we  fitted the flux-density ratios between the NE and SW images using the XX and YY visibilities, as given prior to the calibration of the instrumental polarization. Hereafter, we call $\rho_X$ the flux density ratio obtained from the XX visibilities and $\rho_Y$, that from the YY visibilities. In these fits, we divided each scan into 15 segments of equal duration to perform an  intra-observation analysis similar to that described in the previous section. The time variation of $\rho_{X}$ and $\rho_{Y}$ encodes information about the differential polarization between the two lensed images \citep{mar16,mar15}. The evolution of the differential-polarimetry ratio, which we define as $R_{pol} = (\rho_X/\rho_Y-1)/2$, as a function of the source rotation (as seen from the antenna feeds, which are rotated by $\beta = 45^\circ$ with respect to the horizontal coordinates in Band 6; see \citealt{mar16} for more details), is shown in Fig.\,\ref{fig:diffpol}. As expected due to Earth rotation, $R_{pol}$ follows a sinusoidal behavior, with an amplitude and phase directly related to the polarization of NE and SW. In the same figure we plot the expected value of $R_{pol}$ derived from Eq. 11 of \cite{mar16}, which is the exact expression for differential polarimetry, using the Stokes parameters for NE and SW derived from the full-polarization image. There is  good agreement between the dual-polarization measurements and the full-polarization prediction, although the fit improves slightly (a decrease of 10--15\% in the $\chi^2$) by rotating the SW image by 2$-$5 degrees. This offset may be related to the intra-observation EVPA variability discussed in Sect. \ref{FullPolResSec}, which was smoothed out in the process of image deconvolution.

Regarding the frequency dependence of $R_{pol}$, we do not detect any significant difference among the four spectral windows at a level larger than 1$\sigma$ (see Fig.\,\ref{fig:diffpol}). We note that this differs from the results found by \cite{mar15} toward the same source, where large changes in $R_{pol}$ were seen among the different spectral windows, an effect directly related to a high rotation measure. Our new results indicate that the physical conditions in the magneto-ionic medium that caused the strong Faraday rotation seen in year 2013 changed dramatically between then and our 2019 observations. The maximum $R_{pol}$ measured in our 2019 observations ($\sim 0.1$; Fig.\,\ref{fig:diffpol}) is much higher than the value in 2013 (a few times $10^{-3}$; see \citealt{mar15}). The low $R_{pol}$ values for year 2013 could indeed be caused by Faraday depolarization, directly related to the high Faraday rotation reported for the same data \citep{mar15}.

\section{Conclusions} 

We present ALMA full-polarization observations of the lensed blazar \PKS1830, obtained close in time to the peak of its record-breaking flaring activity in spring 2019. Our results can be summarized as follows:

\begin{itemize}
\item  The two lensed images (NE and SW) show a remarkable difference in polarization. The NE fractional polarization is a factor of $\sim$3 weaker than that of SW. Since the SW image is delayed in time by $\sim$26 days with respect to the NE, this polarization difference implies a significant suppression of the polarization intensity during the flaring activity.

\item The differential electric vector position angle varied at a rate of about 2~degrees per hour during the time span of $\sim 2$~h of our observations.

\item Simultaneously, the relative fractional polarization changed by $\sim 10$\%.

\item Our observations thus show significant variations on short timescales on the order of one hour.

\item The changes in polarization properties are consistent with predictions from models of magnetic turbulence in the jet.

\item The full-polarization calibration allows us to validate the technique of differential polarimetry.

\end{itemize}

It is now possible, with an instrument like ALMA, to follow short timescale variations in the flux and polarization evolution of blazars at millimeter and submillimeter wavelengths, allowing us to obtain  valuable information, such as EVPA and fractional polarization variations on short timescales, to constrain the mechanisms at work in the launching and shaping of jets from active galactic nuclei.

\begin{acknowledgement}
We thank the referee for the constructive and helpful comments to improve the paper. IMV and AMM thank the Generalitat Valenciana for funding, in the frame of the GenT Project CIDEGENT/2018/021. APM acknowledges funding support from National Science Foundation grant AST-1615796. I.A. acknowledges support by a Ram\'on y Cajal grant (RYC-2013-14511) of the ``Ministerio de Ciencia e Innovaci\'on (MICINN)'' of Spain. He also acknowledges financial support from MCINN through the ``Center of Excellence Severo Ochoa'' award for the Instituto de Astrof\'isica de Andaluc\'ia-CSIC (SEV-2017-0709) and through grant AYA2016-80889-P. This paper makes use of the following ALMA data: ADS/JAO.ALMA\#2018.1.00692.S. ALMA is a partnership of ESO (representing its member states), NSF (USA) and NINS (Japan), together with NRC (Canada) and NSC and ASIAA (Taiwan) and KASI (Republic of Korea), in cooperation with the Republic of Chile. The Joint ALMA Observatory is operated by ESO, AUI/NRAO and NAOJ.
  This research has made use of NASA's Astrophysics Data System.
\end{acknowledgement}

\begin{appendix}

\section{Faraday rotation}\label{appendix:Faraday}

\begin{figure}[h] 
\begin{center}
\includegraphics[width=8.5cm]{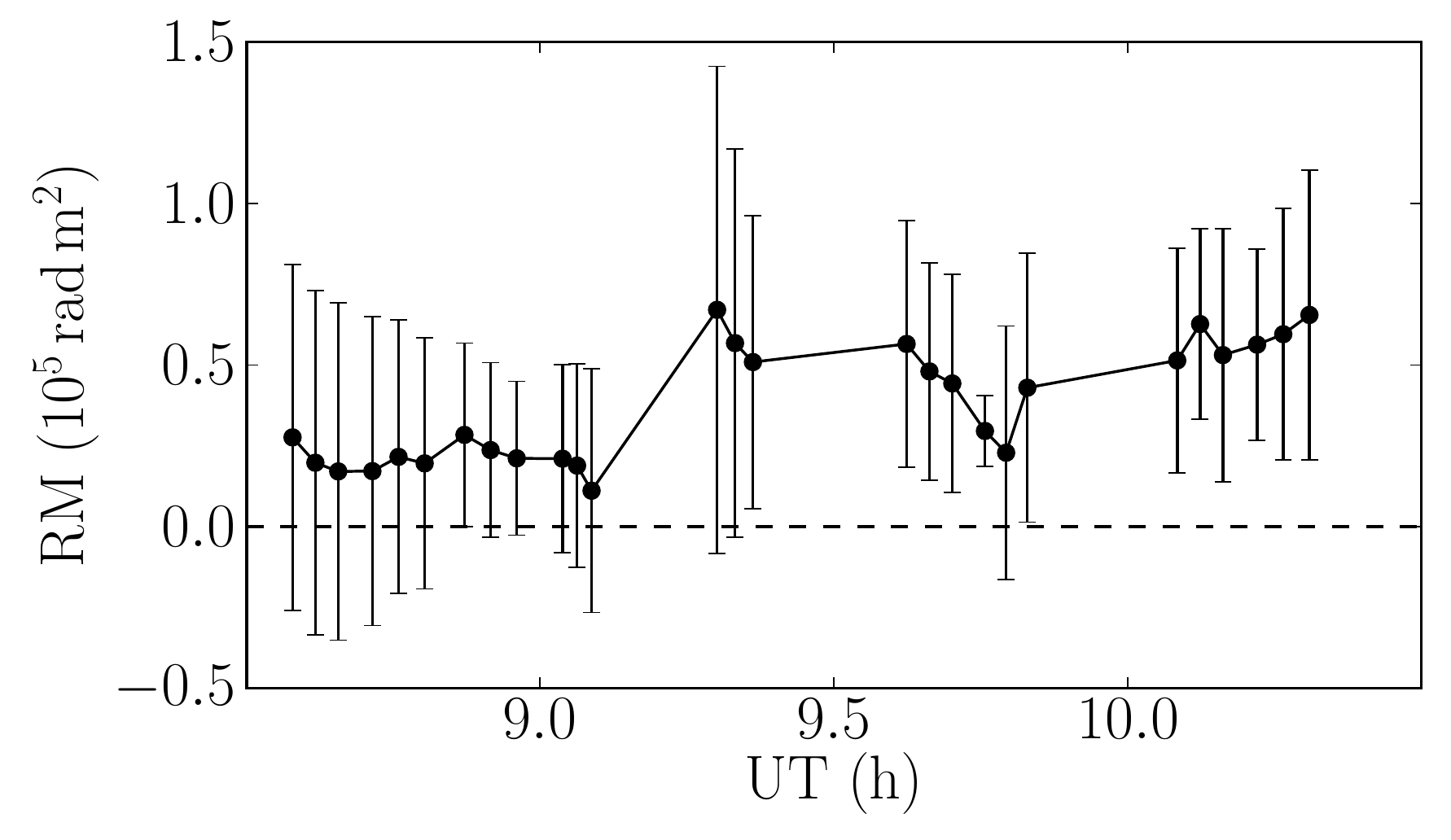}
\caption{Rotation measure at each time interval, fitted with an ordinary linear regression of EVPA difference vs. $\lambda^2$.}
\label{fig:RM}
\end{center} \end{figure}

The EVPA difference between NE and SW   clearly varies in time during the experiment, in a way that slightly depends on frequency (see Fig.\,\ref{fig:curves}A). This intricate behavior may point to a non-negligible contribution from a complicated Faraday screen and/or to internal Faraday rotation in the jet (which would introduce a spectral dependence of the EVPA that would not fit the standard $\phi \propto \lambda^2$ relation). Another possibility could be the presence of residual spectral instrumental effects after the ALMA calibration, although we are not aware of any calibration artifacts that would affect the two lensed images in different ways. In any case, we   used a model of thin Faraday screen (i.e., $\phi \propto \lambda^2$ law) to fit the EVPA difference between the two images at each time bin. The results are shown in Fig.\,\ref{fig:RM}. All the fitted values are compatible with zero at a 2$\sigma$ level, with upper limits (in absolute value) of a few times $10^4$~rad\,m$^{-2}$. We note though that the fitted rotation measure (RM) values are not spread randomly around zero since they are all   positive. This may be indicative of a tentative RM detection in the data. The weighted average of all the RM measurements is $(3.4 \pm 1.7)\times 10^4$~rad\,m$^{-2}$.

\end{appendix}
\end{document}